# Direct Numerical Simulation of High Prandtl Number Fluid Flow in the Downcomer of an Advanced Reactor

**Tri Nguyen, Elia Merzari**
Ken and Mary Alice Lindquist Department of Nuclear Engineering
The Pennsylvania State University, 228 Hallowell, University Park, PA 16802, USA
nguyen.tri@psu.edu; ebm5351@psu.edu

**Cheng-Kai Tai, Igor A. Bolotnov**
North Carolina State University
Department of Nuclear Engineering
ctai2@ncsu.edu, igor_bolotnov@ncsu.edu

**ABSTRACT**

The passive safety is a crucial feature of advanced nuclear reactor (Gen IV) design. During loss of power scenarios, the downcomer plays a crucial role. The fluid-flow behavior in the downcomer can involve forced to mixed to natural convection and characterizing the heat transfer for these changing regimes is a daunting challenge. High-resolution heat transfer numerical database can potentially support the development of precise and affordable reduced resolution heat transfer models. These models can be designed based on a multiscale hierarchy developed as part of the recently DOE-funded center of excellence for thermal-fluids applications in nuclear energy. In this paper, the downcomer is simplified to heated parallel plates and High Prandtl number fluid (FLiBe) is considered for all simulations. The calculations are performed for a wide range of Richardson number from 0 to 400 at 2 different FLiBe's Prandtl numbers (12 and 24) which result in 40 simulated cases in total. Time averaged and time series statistics as well as Nusselt number correlations are investigated to illuminate the mixed convection behaviors. The calculated database will be instrumental in understanding the flow behaviors in the downcomer. Ultimately, we aimed to evaluate existing heat transfer correlations and some modifications will be proposed for cases where no satisfactory choice is available.

## 1. Introduction

Molten Salts are promising coolants for advanced reactor designs thanks to their special thermophysical and neutronic properties [1]. For example, FLiBe, a mixture of lithium fluoride (LiF) and beryllium fluoride ($BeF_2$), is one of the most well studied molten salts, as it was successfully demonstrated in the Molten Salt Reactor Experiment (MSRE) [2] from 1965 to 1969 at Oak Ridge National Laboratory. However, a comprehensive heat transfer characteristics database for molten salts is still not yet established as traditional water-based correlations (e.g., Prandtl unity heat transfer correlations) may not be applicable. The turbulence and heat transfer modelling for such molten salts is very challenging, and there is a lack of DNS data that can be used to advance reactor design. Currently, the experimental results and high-fidelity numerical simulations are ineffectively applied to complement low resolution methods. As a result, the Center of Excellence for Thermal Fluids Applications in Nuclear Energy [3] has been established to develop and validate trustworthy and affordable advanced thermal-hydraulic models. The mixed convection effects in the downcomer of advanced reactor have been identified as a part of a challenge problem driven by industrial needs. The generated database, driven by requirements identified by our industry partners, will be used to illuminate the flow behaviors in the downcomer, and finally, it will be used to develop reduced resolution models and evaluate existing heat transfer correlations.

In this paper, we will discuss Direct Numerical Simulation (DNS) performed to investigate the mixed convection heat transfer in the downcomer for a wide range of Richardson number from 0 to 400 at 2 different FLiBe's Prandtl (Pr) numbers (12 and 24). The Spectral Element code NekRS [4] is used for all simulation cases thanks to its superior calculation speed. NekRS is a GPU version of Nek5000 [5] which has demonstrated its capability to perform high fidelity LES and DNS in both simple and complex geometries [6,7,8]. The downcomer can be simplified as flow between two heated parallel plates. The working fluid is FLiBe which is the interest to our industrial partner Kairos Power. Forced convection cases data has been compared with the available DNS data of Kasagi et.al [9,10,11] and good agreements has been achieved. Noted that for high Pr fluid, the momentum diffusivity dominates the thermal diffusivity, thus the Bachelor length scale must be applied to assess the computational mesh resolution requirements. More details can be found in [12].

The main difference between performing low-Pr and High-Pr DNS is the high computational cost coming from the necessity of resolving Batchelor length scale which can be orders of magnitude higher than for low Pr fluids. The mesh sizes contrast has been demonstrated for high-Pr (FLiBe) [12] and low-Pr fluid [13] in the DNS of the parallel plates. To overcome the computational cost of DNS, Reynolds Averaged Navier Stokes (RANS) turbulence models can be used. However, this can present notable uncertainty due to the restrictive and limiting assumptions of the RANS formulation. The intrinsic limitations in RANS models lead to the necessity to develop with DNS datasets [14,15] to benchmark their effectiveness. The DNS database in this work will be generated for parallel plates with the focus on differential heating/cooling applied to case 1 and symmetric cooling applied to case 2. The computational setups will be described in detail in section 2. Section 3 will focus on the results and discussion where the calculated time averaged statistics such as Turbulence Kinetics Energy (TKE) and Turbulence Heat Flux (THF) budgets have been analyzed. Furthermore, snapshots Proper Orthogonal Decomposition (POD) introduced by Sirovich [16] has been implemented for better understanding and interpreting the flow behaviors. Time signal and frequency analyses have also been considered, and differences in Power Spectral Density (PSD) behaviors between cases 1 and 2 have been revealed. Moreover, the Nusselt number calculations have been estimated for each case and trends have been compared against existing experimental data. This analysis shows a good agreement between numerically estimated Nusselt numbers and available correlations as well as experimental data. We are also proposing a novel correlation for high Pr flows as they are transitioning from mixed convection to natural convection. Section 4 summarizes the paper and provides some conclusions.

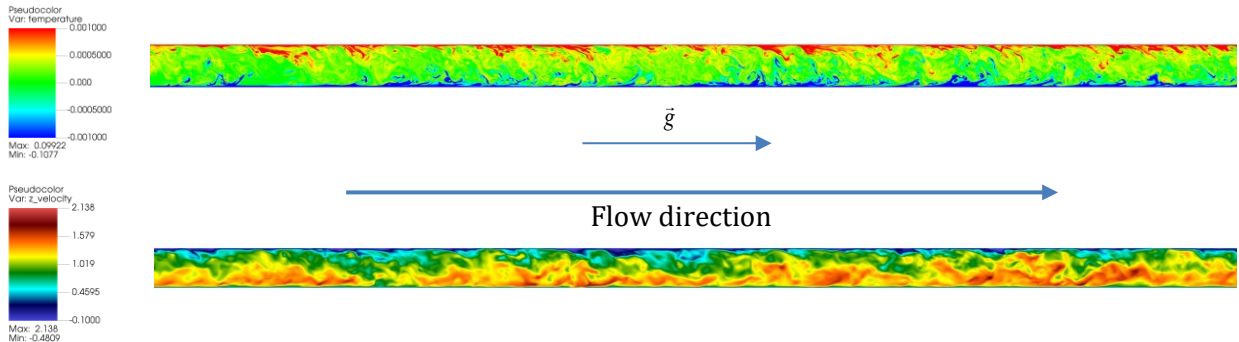

**Figure 1.** An example of velocity and temperature fields of downcomer (parallel plates) with mix convection for case 1 at Pr = 24.

## 2. Numerical Methodology

### 2.1. Numerical Method

The highly scalable high-order spectral element code NekRS is chosen for both LES and DNS thanks to its capability to perform high fidelity LES and DNS for not merely canonical but also complex geometries problems [12,17,18] at superior calculation speed compared to its predecessor Nek5000. NekRS takes advantage of all features in Nek5000 and utilize them effectively in GPU architectures, thus drastically improving the calculation speed compared to Nek5000. In NekRS, the GPU kernels are written in OCCA library to condense between different parallel languages such as OpenCL, CUDA, and HIP. Users can use OCCA to implement the parallel kernel code in C++ language, OKL to convert Nek5000 cases to NekRS ones. It is important to note that Nek5000 and NekRS can address cases with variable thermophysical properties using low-Mach number approximation [4,19]. The main difference between low-Mach compressible Navier-Stokes (NS) equations and incompressible NS equations is the non-zero divergence term in the continuity equation which accounts for the change in density. However, in this work, we only considered incompressible NS equations with constant properties Newtonian fluid (FLiBe) subject to buoyancy, modeled though the Boussinesq approximation:

$$\nabla \cdot u = 0 \tag{1}$$

$$\rho \left( \frac{\partial u}{\partial t} + u \cdot \nabla u \right) = -\nabla P + \mu \nabla \cdot \nabla u + \rho g (1 - \beta(T - T_0)) \tag{2}$$

$$\rho c_p \left( \frac{\partial T}{\partial t} + u \cdot \nabla T \right) = \nabla \cdot (k \nabla T) \tag{3}$$

where Eq. (1) is mass continuity, Eq. (2) holds momentum conservation and Eq. (3) is the energy equation represented in terms of temperature. u and g are vector velocity and gravity, respectively, $\rho$ is the density of FLiBe, $\mu$ represents its dynamic viscosity, $\beta$ represents the heat expansion coefficient, k is the conductivity and cp the heat capacity. The NS equations were solved in NekRS using the spectral element method (SEM) introduced by Patera [20]. It is a Galerkin-type method that delivers good geometric flexibility like finite element methods, with small numerical dispersion and rapid convergence properties typical of spectral methods [4]. In SEM, the computational domain is discretized into E hexahedral elements and represents the solution as a tensor-product of $N^{th}$-order Lagrange polynomials based on Gauss-Lobatto-Legendre (GLL) nodal points, which leads to $E(N+1)^3$ degrees of freedom per scalar field. GLL points are selected for efficient quadrature for both velocity and pressure spaces. The pressure is solved at the same polynomial order of the velocity N ($P_N$ -$P_N$ formulation).

We simulate 40 cases in this work. The nondimensional form of NS equations (1), (2), (3) is applied:

$$\nabla \cdot u^* = 0 \tag{4}$$

$$\left( \frac{\partial u^*}{\partial t^*} + u^* \cdot \nabla u^* \right) = -\nabla P^* + \nabla \cdot \frac{1}{Re} \nabla u^* - Ri * T^* \tag{5}$$

$$\rho^* \left( \frac{\partial T^*}{\partial t^*} + u^* \cdot \nabla T^* \right) = \nabla \cdot \frac{1}{Pe} \nabla T^* \tag{6}$$

Where $u^*$, $t^*$, $\rho^*$, $P^*$, $T^*$, Re, Ri, Pe are nondimensional velocity, time, density, pressure, temperature, Reynolds number, Richardson number and Peclet number, respectively. $u^*$=u/U, U is the inlet bulk velocity, $t^*$ = t/(D/U), D = 2$\delta$, D is the hydraulic diameter, $\delta$ is the half channel heigh, $P^* = P/(\rho_0 U^2)$, $\rho_0$ is the inlet density, $\rho^* = \rho/\rho_0 = 1$; $T^* = (T-T_0)/\Delta T$, $T_0$ is the inlet temperature, $\Delta T$ is the temperature difference between inlet and outlet (for case 2 in Figure 2 B) or between the 2 walls (for case 1 in Figure 2 B). Note that $Re = (DU\rho_0)/\mu$, $Fr = U/\sqrt{gD}$, $Pe = PrRe = (\rho_0 UD c_{p0})/k$, $Ri = \beta\Delta T/Fr^2$, where $Fr$ is the Froude number.

To characterize the mix convection phenomena, precisely, the relative strength between forced and natural convection, The Richardson number is introduced. The Richardson number $Ri$ defined quantitatively the force term in equation (5), and it is the ratio of Grashof number to the square of $Re$ number, as shown in (7) [12,21]:

$$Ri = \frac{Gr}{Re^2} = \frac{\beta\Delta T}{Fr^2} \tag{7}$$

## 2.2 Numerical setup

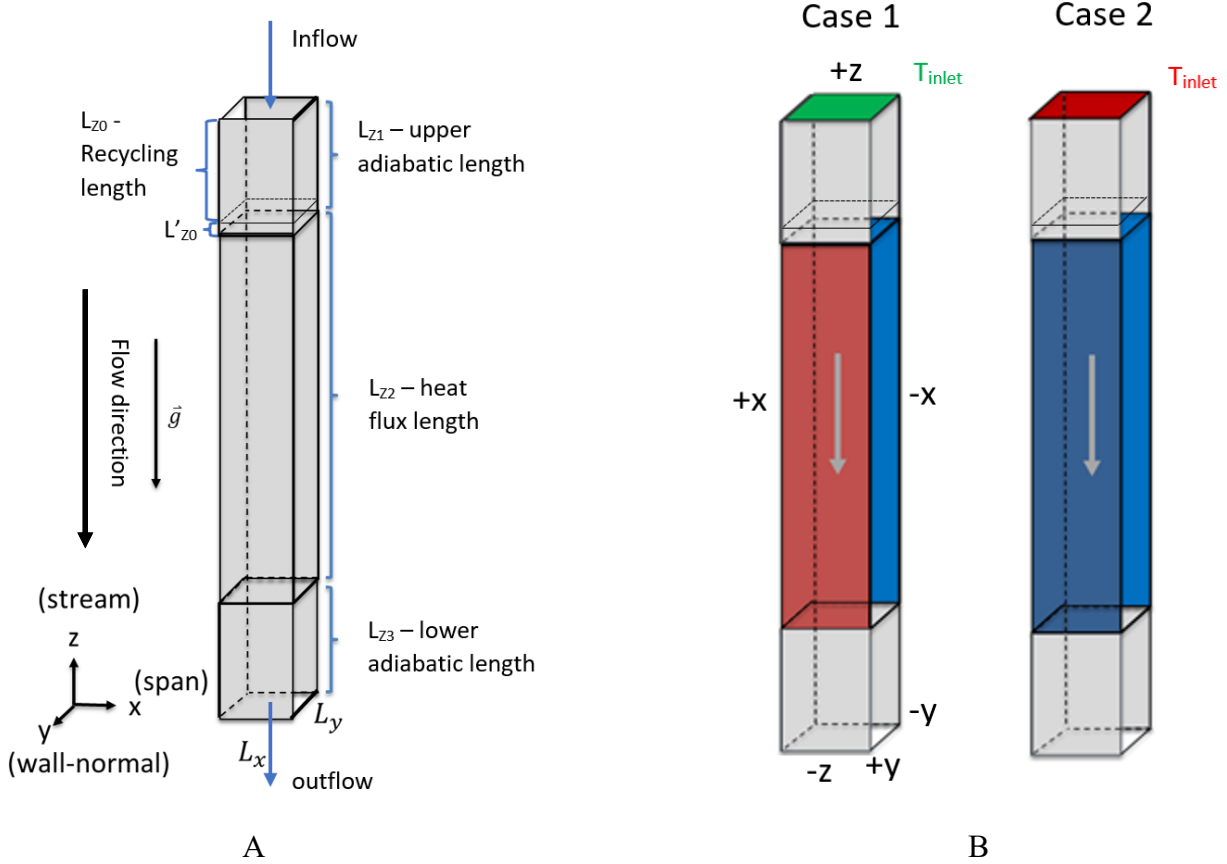

**Figure 2.** A – The simplified downcomer presented as parallel plates computational box. B – Boundary condition map for investigated cases.

The computational model for the parallel plates case is shown in Figure 2A. The fluid domain is divided into 3 sections: upper adiabatic, heat flux and lower adiabatic. The ratio between them, as well as the spanwise $L_x$ and wall normal $L_y$ lengths are defined according to the interest to our industrial partner Kairos Power.

$$L_x = \pi \, L_y \tag{8}$$

$$L_{Z0} = L_{Z3} = 10L_y \qquad (9)$$

$$L_{Z1} = L_{Z0} + L'_{Z0} = 10L_y + 2L_y = 12L_y \qquad (10)$$

$$L_{Z2} = 100L_y \qquad (11)$$

According to (8), (9), (10), and (11), the streamwise length of the channel is $122L_y$. A fully developed inflow condition to the head flux section at desired Re are generated by using the recycling method. Note that for upper adiabatic length, $L_{Z1} = L_{Z0} + L'_{Z0}$ where $L'_{Z0}$ is the distancing between recycling length and heat flux length, which equals to 1D. $L'_{Z0}$ is introduced, and the gravity is not applied in the upper adiabatic length to ensure that the fully developed turbulence flow generated by recycling entered the heated length properly. Gmsh [22] has been used to generate the mesh. The resulting mesh has 250000 hexahedra. All simulation cases use the same mesh but different polynomial orders. For instant, case 1 at Re = 10000 with Pr = 12 required 11th polynomial order, which results in 677 million degrees of freedom. The details of mesh sizes corresponding to each case can be found in table 2. The maximum $y^+$ are lower than 0.1 along the walls and the gravitational direction was chosen to be on the z axis, as shown in Figure 2A. A complete discussion of mesh convergence is available in [12].

We investigated two classes of cases (case 1 and case 2) as shown in Figure 2B, representing different cooling conditions at the wall boundary. The working fluid is FLiBe at 695 $^0$C and 560 $^0$C which corresponds to Pr = 12 and 24 in nondimensional terms. Neuman boundary condition (BC) imposing constant heat flux is applied for two opposite vertical walls in +y and -y direction. The spanwise directions +x and -x are periodic. The Dirichlet BCs imposing constant temperature are used at inlet. The detail discussions of the temperature difference ΔT, and the nondimensional heat flux and Ri number for case 1 and case 2, are provided in [12,21]. The boundary conditions (BCs) for case 1 and case 2 are summarized in table 1.

The boundary conditions (BCs) for case 1 and case 2 are summarized in table 1.

| BCs Geometry | Velocity BCs | | temperature BCs | |
|---|---|---|---|---|
| | case 1 | case 2 | case 1 | case 2 |
| +x and -x | Periodic | Periodic | Periodic | Periodic |
| +y, for $L_{Z1}$ and $L_{Z3}$ | No-slip | No-slip | Isolated | Isolated |
| +y, for $L_{Z2}$ | No-slip | No-slip | Heat flux | Heat flux |
| -y, for $L_{Z1}$ and $L_{Z3}$ | No-slip | No-slip | Isolated | Isolated |
| -y, for $L_{Z2}$ | No-slip | No-slip | Heat flux | Heat flux |
| +z | inflow | inflow | $T^* = 0$ | $T^* = 0$ |
| -z | Pressure outflow | Pressure outflow | Isolated | Isolated |

Table 1: the velocity and temperature BCs applied for case 1 and case 2

The initial conditions are the same for all cases,

$$T^* = 0 \qquad (12)$$

$$u_x^* = u_y^* = u_z^* = 0 \qquad (13)$$

The summary of simulations cases and the corresponding meshes resolution are reported in Table 2. Ri = 0 means forced convection cases. Each Re number has 4 corresponding Ri number and 2 Pr numbers, which lead to 8 simulation cases applied for case 1 and case 2. cases at Re = 10000 and Pr = 24 are not being performed due to the limitation in computational resources. The conditions listed have been driven by input from our industry partner Kairos Power, and they are relevant to FHR accident and operational conditions.

| Re | cases | | | | Mesh Resolution | | | |
|---|---|---|---|---|---|---|---|---|
| | case 1 | | case 2 | | Pr =12 | | Pr =24 | |
| | Ri | Gr | Ri | Gr | Polynomial order | Degree of freedom | Polynomial order | Degree of freedom |
| 5000 | 0 | 0 | 0 | 0 | 7 | 172 million | 9 | 365 million |
| 5000 | 8 | $2\times10^8$ | 0.04 | $10^6$ | | | | |
| 5000 | 80 | $2\times10^9$ | 0.4 | $10^7$ | | | | |
| 5000 | 400 | $10^{10}$ | 2 | $5\times10^7$ | | | | |
| 7500 | 0 | 0 | 0 | 0 | 9 | 365 million | 12 | 866 million |
| 7500 | 3.56 | $2\times10^8$ | 0.0178 | $10^6$ | | | | |
| 7500 | 35.56 | $2\times10^9$ | 0.178 | $10^7$ | | | | |
| 7500 | 177.78 | $10^{10}$ | 0.889 | $5\times10^7$ | | | | |
| 10000 | 0 | 0 | 0 | 0 | 11 | 677 million | Not performed | Not performed |
| 10000 | 2 | $2\times10^8$ | 0.01 | $10^6$ | | | | |
| 10000 | 20 | $2\times10^9$ | 0.1 | $10^7$ | | | | |
| 10000 | 100 | $10^{10}$ | 0.5 | $5\times10^7$ | | | | |

Table 2: the simulation cases with their corresponding Reynold (Re), Grashof (Gr), Richardson (Ri) numbers and meshes resolution.

The advantage of using NekRS on GPUs over Nek5000 has been evaluated by comparing the computing time for 1000 timestep for case 1 at Re = 1000 and Ri = 400 on the Summit supercomputer using 5 compute nodes. Nek5000 completed 1000 timestep in around 1800 s (≈1.8s per timestep) on 120 IBM POWER9 CPU cores, whereas NekRS completed 1000 timestep in just 20s on 30 NVIDIA Volta V100 GPUs (≈0.02s per timestep), demonstrating superior calculation speed toward Nek5000.

## 3. Results and Discussion

### 3.1. Validation

Due to the lack of relevant references for the simulation conditions of mixed convection cases (Ri ≠ 0) as referenced in table 2, a separate simulation campaign with forced convection case (Ri = 0) has been performed to compare with Kasagi et.al [9,10,11] DNS data. The Re and Pr have been selected to match with Kasagi et.al cases. The averaged streamwise velocity and RMS fluctuation, as well as TKE and Turbulence Heat Flux (THF) budgets, are compared with previous DNS data, and excellent agreements have been achieved. The comparison results are delivered in detail in [12, 21, 23]. Noted that the agreement with Kasagi DNS data could only be archived with specific runtime settings, which are then implemented for all cases in table 2.

### 3.2. Time-averaged statistics of heated parallel plates at different Richardson, Reynolds, and Prandtl numbers.

Before collecting data, all cases have been run for at least 3 flows through time to reach statistical convergence, as recommended by runtime settings acquired from our validation campaign. The effect of buoyancy for cases with Ri number ranging from 0.04 to 400 at different Re (5000, 7500, 10000) and Pr (12 and 24) numbers is investigated and analyzed. A series of comparison campaigns have been conducted for all cases including normalized averaged and RMS of velocity and temperature, TKE, THF. All the turbulence data presented in this section has been collected on a z-plane cross-section at z = $L_{Z1}$ + 0.5$L_{Z2}$ = 62Ly, where the flow is considered fully developed. Noted that for case 1, the turbulence statistics including average, rms, TKE and THF budgets of velocity and temperature were collected for both hot and cold wall due to the asymmetric temperature BCs. For case 2, because of the symmetric temperature BCs, the statistics are the same for both sides of wall. We also compared the turbulence statistics results for case 1 at Re = 5000, Pr = 12 at different spanwise length ($L_x$ = $\pi/2$, $\pi$, and $2\pi$) and no changes in turbulence statistics has been observed.

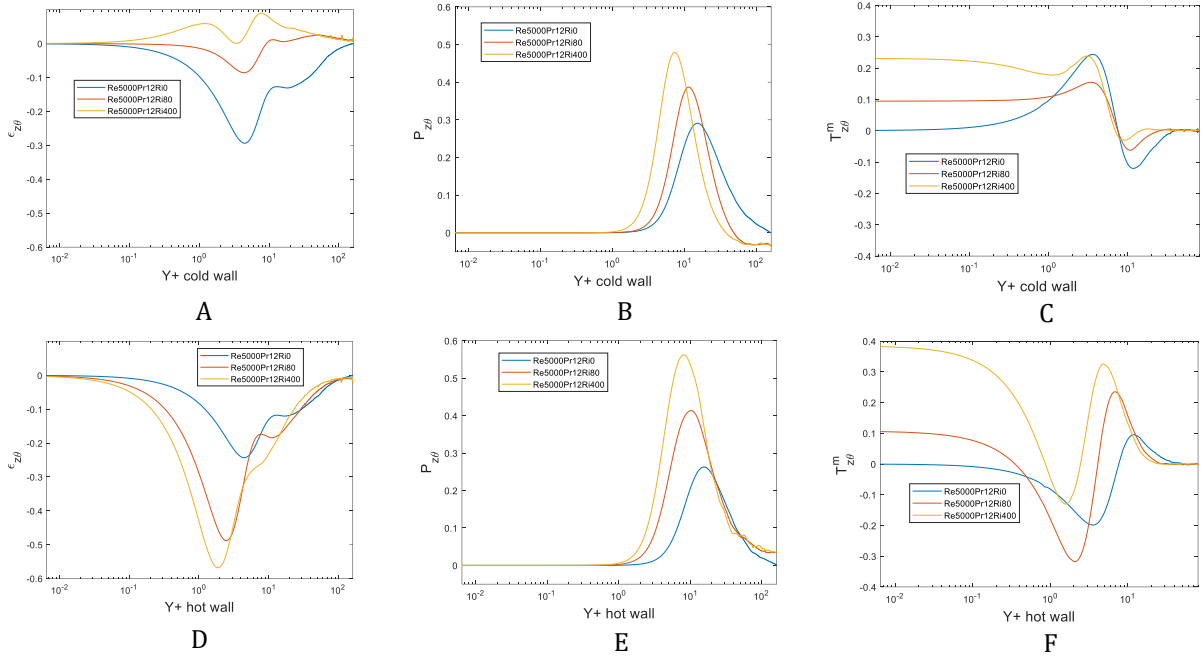

**Figure 3:** THF budgets for cases 1 at Re = 5000, Pr = 12. A, D – Streamwise THF dissipation ($\epsilon_{z\theta}$) at A - cooled wall, D - heated wall. B, E – Streamwise THF production ($P_{z\theta}$) at B - cooled wall, E - heated wall. C, F – Streamwise THF viscos diffusion ($T_{z\theta}^m$) at C - cooled wall, F - heated wall.

The characteristics of the average and rms of velocity and temperature profile, as well as the TKE budgets of case 1 and case 2 at Re = 5000 have been discussed in detail in [12]. In general, the behavior of cases at Re = 7500 and 10000 are similar to case at Re =5000 as the Ri number increases. It can be seen in Figure 3 that the THF budgets are strongly alter as the Ri number changes [12]. We observe a negative THF production region as shown in Figure 3B, which is caused by flow acceleration due to the buoyancy force. However, unlike the negative TKE production that increase strongly as Ri increase, the THF accounts for the intensity of temperature to its calculation, which results in very small changes as Ri number increase.

Figure 4 shows the turbulence statistics behavior of several cases at the same Re but different Pr numbers. At the same Re number, the larger the Pr number, the weaker the Ri number effect. Overall, the amplitude of turbulence statistics decreases as the Pr number increases. In some conditions, for instance case 1 at Ri = 80, the flow shifts from a mixed convection case at Pr = 12 to a likely forced convection case at Pr = 24, the negative production region vanishes as shown in Figure 4B and 4E. For case 2, the Pr effect strongly reduces the TKE production double peaks amplitude, thus reducing the intensity of the mixed convection effect. The bigger Ri is, the stronger this effect is, as shown in Figure 4C and 4F.

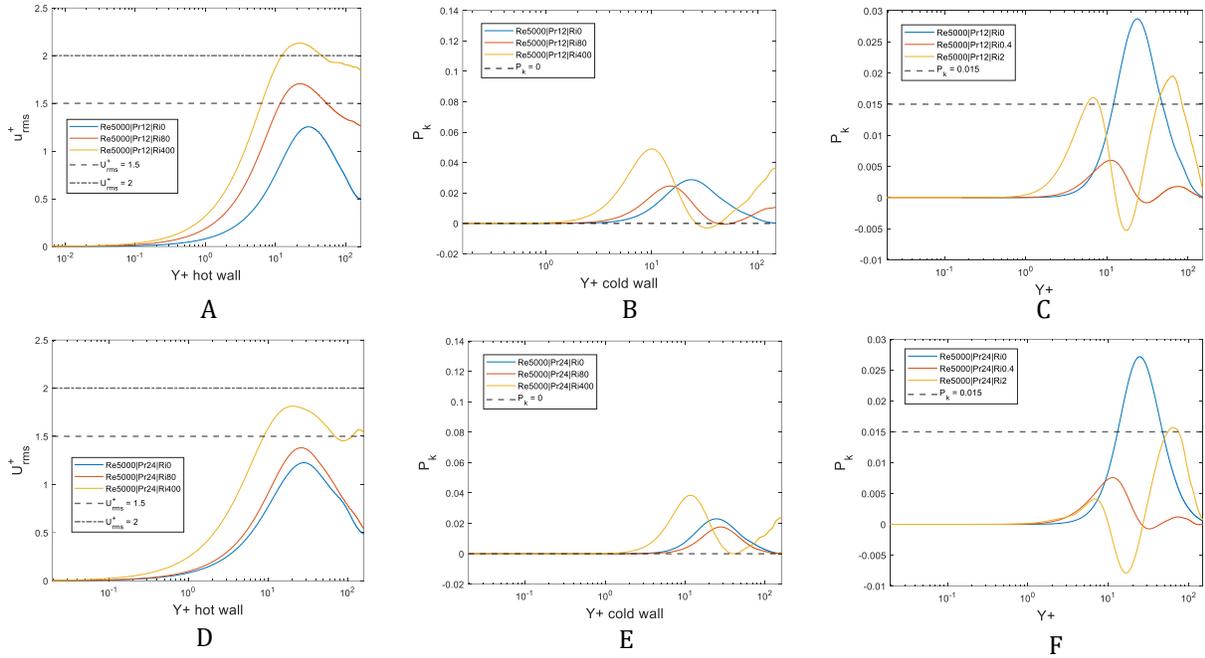

**Figure 4:** Turbulence statistic comparison between cases at Pr = 12 and 24 with Re = 5000. A, D – RMS streamwise velocity $u^+_{rm}$ for case 1 at hot wall. B, E – TKE production $P_k$ for cases 1 at cold wall. C, F – TKE production for cases 2.

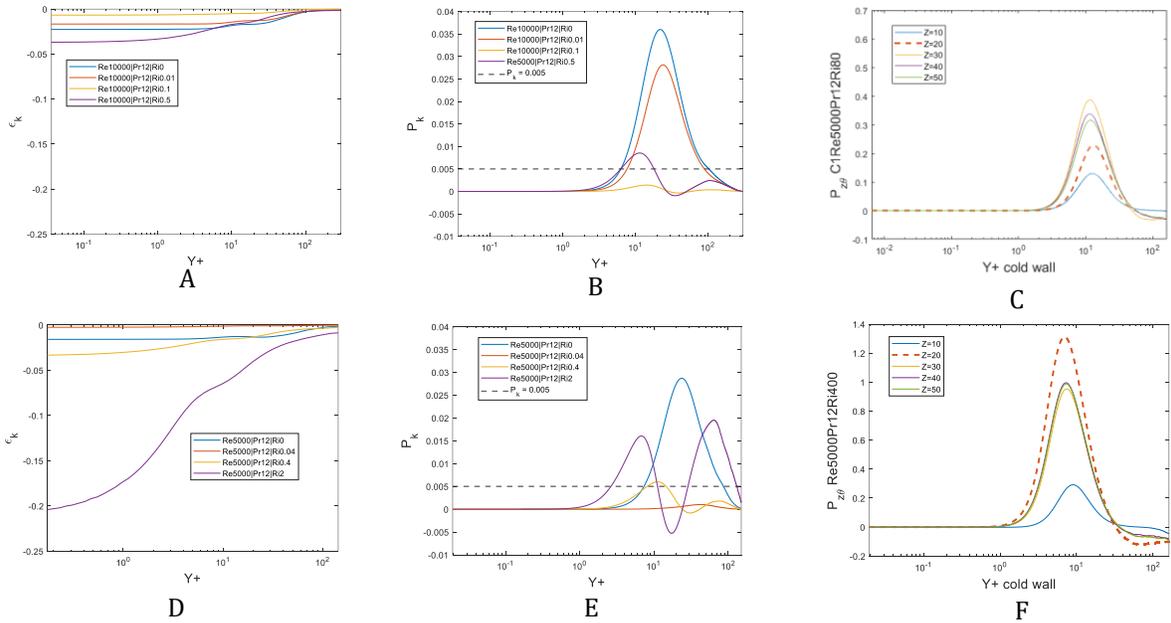

**Figure 5:** A, B, D, E – TKE budgets comparison between cases 2 at Pr = 12 with Re = 5000 and 10000. A, D - TKE dissipation $\epsilon_k$. B, E - TKE Production $P_k$. C and F - THF production of cases 1 at Re = 5000, Pr = 12, Ri = 80 and 400 at multiple streamwise locations on cold wall side.

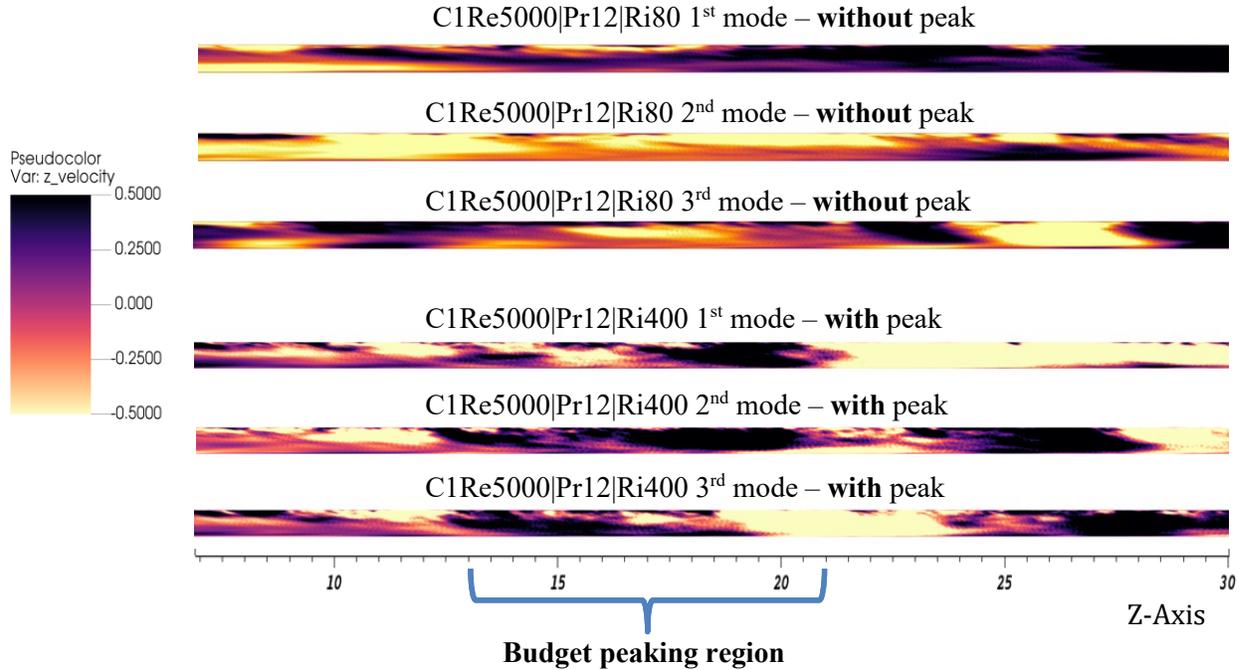

**Figure 6:** streamwise velocity POD of cases 1 at Re = 5000, Pr = 12, Ri = 80 and 400.

Figure 5 (A, B, D, E) shows the TKE budgets behavior of case 2 at different Re but the same Pr number. It is important to note that case1 and 2 in table 2 use the same set of Gr numbers applied for each group of Re number which lead to different set of Ri numbers. For forced convection cases, the higher the Re number is, the stronger the budgets are. That can be explained by the enhancement of turbulence intensity at high Re number. However, while mix convection cases depend strongly on the Ri numbers, Re numbers have little impact on the flow behavior. At the same Ri number, the budgets, as well as the average and rms of velocity behaves similarly, as shown in Figure 5A, 5B, 5D, 5E. We also collected the statistics at different streamwise locations of the channel to evaluate changes across the streamwise direction. The data has been gathered at z = 10, 20, 30, 40, 50 and sample results are shown in Figure 5C and 5F. It is observed that there are budgets peaking at z =20 for several cases, as shown in Figure 5F. To understand the physics behind this phenomenon, POD analyses [7,16] has been implemented for a series of cases with and without budget peaking. The POD is chosen because it us a very effective mathematical method to identify the dominant flow structure, which in turn can reveal the difference in dynamic behavior between cases with and without peaking. To perform POD, 1000 instantaneous snapshots, sufficiently separated in time to avoid excessive correlation, are generated for each simulation case as POD input data. The first 3 POD modes corresponding to sample cases with and without budgets peaking at z = 20 are shown in Figure 6. It can be clearly seen that the POD modes of cases with budget peak present significant differences from the non-peaking counterparts. In the region of interest, i.e., z from 13 to 21, the 1st and 2nd modes of a case at Re = 5000, Pr = 12 and Ri = 400 shows high energy content developing from the hot wall. This effect becomes more marked as z increase from 13 to 21. The size of the energy structure continues to grow until it reaches the cold wall then it decreases in strength. Such a flow structure cannot be seen in cases at Re = 5000, Pr = 12 and Ri = 80 where the budgets peaking is absent. We also observe the same flow behavior for other cases with budgets peak, but we do not present it here in the interest of brevity.

### 3.3. Time series and frequency analysis.

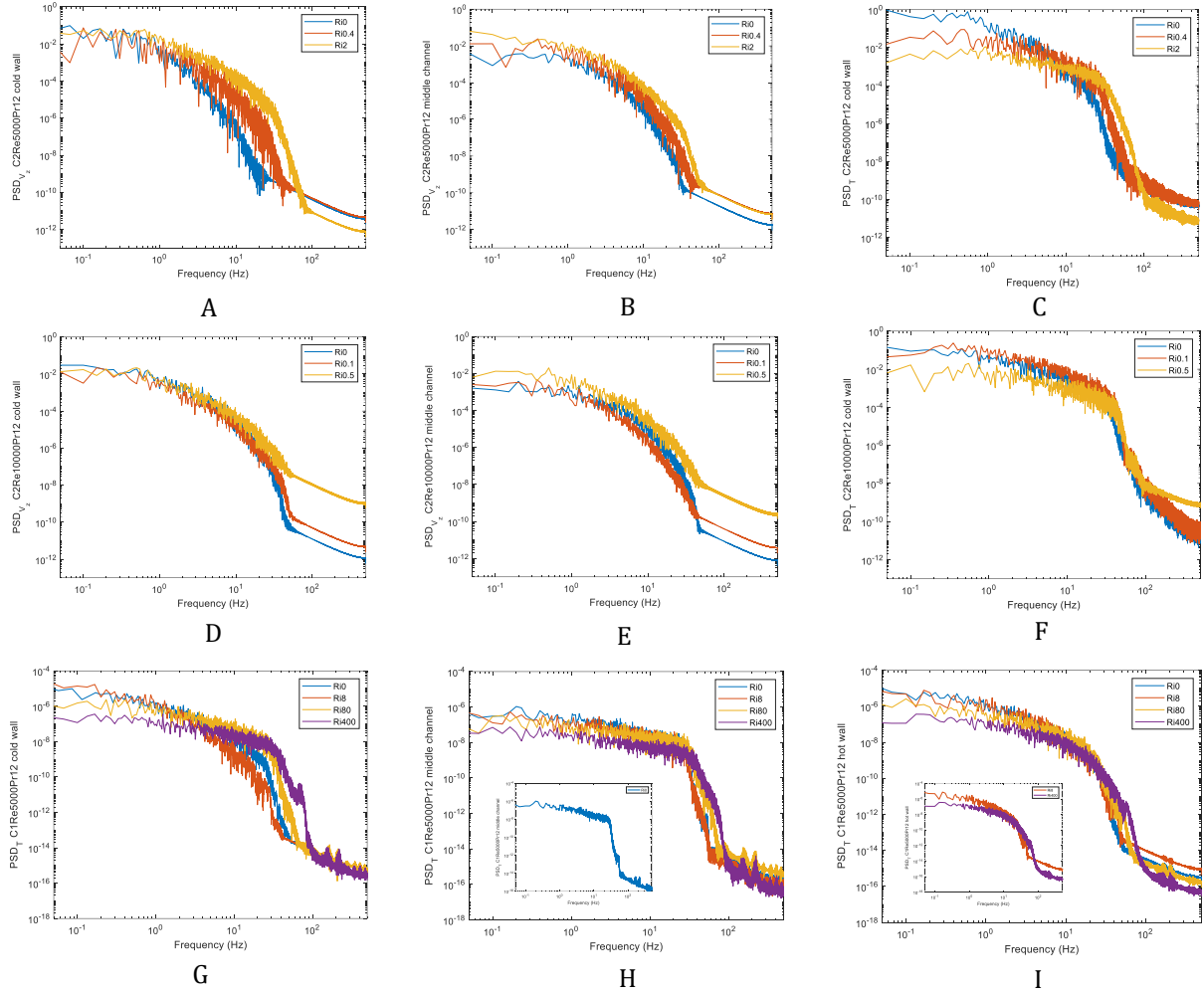

**Figure 7:** A, B, C, D, E, F – PSD of cases 2 at Re = 5000 and Re = 10000 with the same Pr Number (Pr = 12). A, D - PSDs of streamwise velocity for a point near wall. B, E - PSDs of streamwise velocity for a point in middle of the channel. C, F - PSDs of temperature for a point near wall. G, H, I – PSDs of temperature for cases 1

In this section, time signals of streamwise velocity and temperature for cases 1 and 2 at different Re number are considered. The data is collected from points near wall and points in the middle of the channel at z = 30 for a 1.5 flow through time. An overview of the time series behavior can be found in [12,21]. Figure 7 (A, B, C, D, E, F) shows the Power Spectral Density (PSD) of velocity and temperature for case 2 at Re = 5000 and 10000 with the same Pr number. In general, high PSD values are observed at low frequencies, the energy content gradually decreases as the frequency increases. For cases at Re = 5000, it can be recognized in Figure 7C that there are noticeable shifts from low to high frequency as the Ri increase, which corresponds to the evolution from forced to natural convection. The shift at high frequency occurs between 20 Hz and 70 Hz. There are marked differences in PSD between points in the interior and points near wall, likely due to the relative difference in strength of the buoyancy force. Figures 7G, 7H, 7I point to a peculiar peaking of the temperature PSD, with different intensities around a transition region centered around 70 Hz. Based on our observation of temperature and velocity patterns (section 3.2), this peaking could be caused by turbulent structure formation in the early part of channel. We note that the PSD peaking is not present in case 2.

### 3.4. Nusselt number dataset for FLiBe

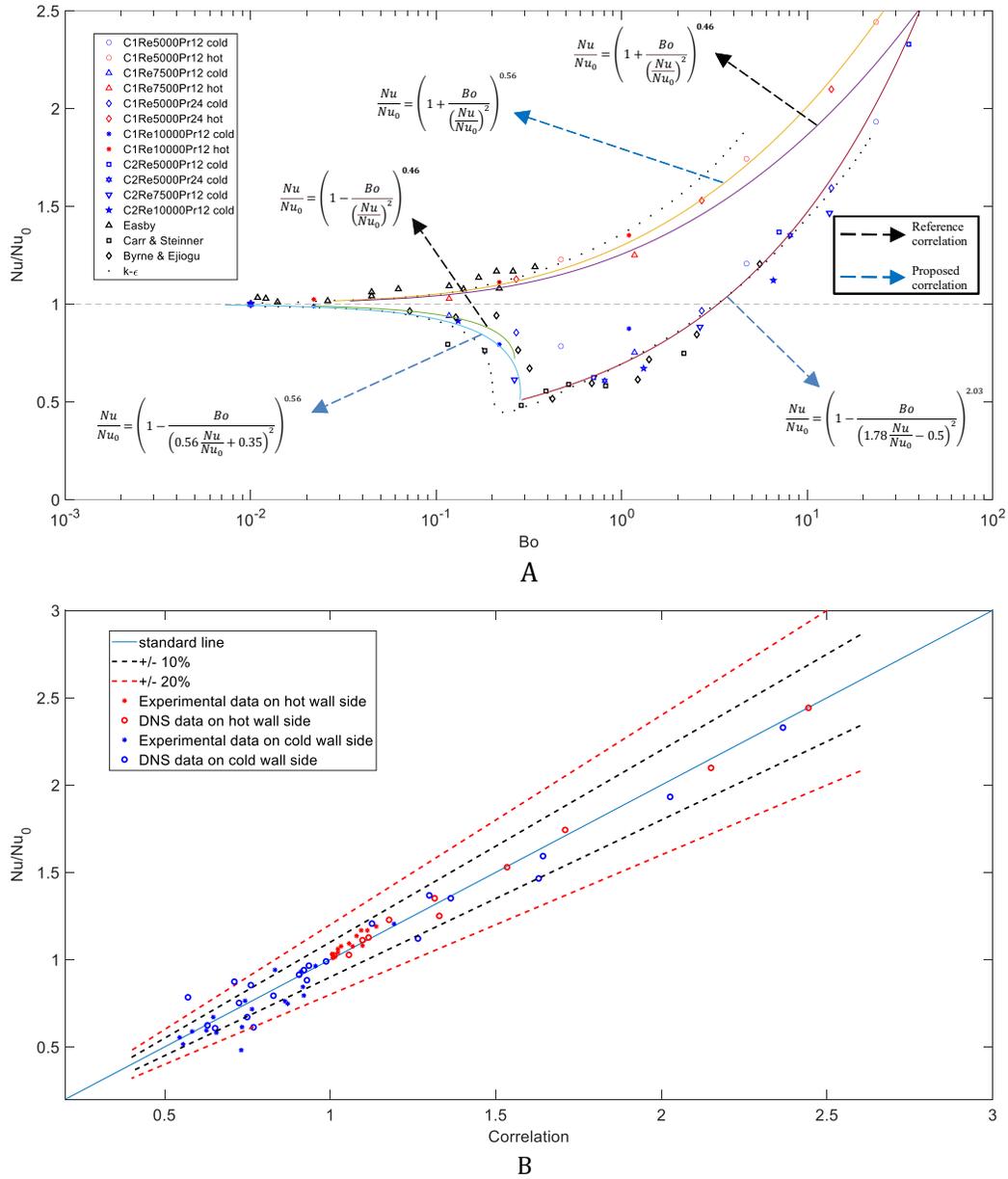

A

B

**Figure 8:** A – Nusselt number dataset of simulation cases in table 2 as a function of Buoyancy number (Bo). Red datapoints indicated the Nu numbers on hot wall side, blue datapoints – Nu numbers on cold wall side, black datapoints – Nu number of reference experiment. B – Nu number line comparison between correlation and experimental & DNS data.

The Nusselt (Nu) number is an important dimensionless number which define the effectiveness of convection heat transfer. For relatively new non-traditional coolant such as molten salt like FLiBe, there is a lack of reliable heat transfer correlations for a broad range of Grashof and Reynolds numbers. In this work, the Nu number has been calculated for each case in table 2 and represents as a function of Buoyancy number (Bo) [24]. Although the local Nu number was obtained across $L_{Z2}$ and $L_{Z3}$ length, for the interest of brevity, the Nu dataset is presented at fully developed region in this work, and it is normalized by forced convection case corresponding to each group of Re number, as shown in Figure 8. The experimental data [25,26,27,28] was collected for air flows in a heated pipe, and a summary of experimental data can be found in the work of Cotton and Jackson [24]. The RANS k-ε results in their work [24] has also been presented for reference. Figure 8A shows the trend of Nu number as Ri number increase from forced to natural convection. cases 1 contributes to Nu data on both hot and cold wall side, whereas cases 2 only represents data on cold wall side.

Overall, the DNS dataset follows the trends of the experimental data. On the hot wall side, the Nu number increase as Ri number grows which can be explained by the opposite direction of the increasing buoyancy force enhancing the turbulence [12]. On the cold wall side, the Nu number initially decrease as the Bo number starts to increase from forced convection (Bo = 0) to Bo = 0.3, forming the heat transfer impairment region. Then the Nu number increases steadily and tend to converge with its counterpart on the hot wall side when Bo is big enough (natural convection). This behavior can be interpreted by the results from section 3.3: starting from Bo = 0 to 0.3, the buoyancy force starts accelerating the flow, the turbulence flow is laminarized (the TKE production becomes zero) and the Nu number in turn declines. Following the growth of Bo, the laminarization effect is maximum at around Bo = 0.3 corresponding to the minimum Nu, and the turbulence is reestablished when Bo > 0.3. After that, the Nu number keeps increasing, the natural convection is established for both hot and cold wall, and there is an increasingly smaller difference in heat transfer coefficient between the two walls.

An effort has been made to find the Nu number correlation describing the trend of current DNS Nu dataset. A reference correlation from [24] has been considered for further development:

$$\frac{Nu}{Nu_0} = \left(1 \pm \frac{Bo}{\left(\frac{Nu}{Nu_0}\right)^2}\right)^{0.46} \tag{14}$$

The implementation of equation (14) is shown in Figure 8A. It can be observed that eq. (14) could not describe the trend of the presented data perfectly. On the cold wall side, it can characterize the dataset behavior in a very limited manner. A generalized Nu correlation has been offered for better representation of the DNS and experimental dataset:

$$\frac{Nu}{Nu_0} = \left(1 \pm \frac{Bo}{\left(a\frac{Nu}{Nu_0} + b\right)^2}\right)^{c} \tag{15}$$

Where a, b, c are the variable parameters. Applying MATLAB Curve Fitting Toolbox [29], the new Nu correlation for the current Nu dataset is found as following:

For the hot wall side:

$$\frac{Nu}{Nu_0} = \left(1 + \frac{Bo}{\left(\frac{Nu}{Nu_0}\right)^2}\right)^{0.56} \tag{16}$$

For the cold wall side:

$$\frac{Nu}{Nu_0} = \left(1 - \frac{Bo}{\left(0.56\frac{Nu}{Nu_0} + 0.35\right)^2}\right)^{0.56} \tag{17}$$

$$\frac{Nu}{Nu_0} = \left(1 - \frac{Bo}{\left(1.78\frac{Nu}{Nu_0} - 0.5\right)^2}\right)^{2.03} \tag{18}$$

Where eq. (17) is applied for Bo ranging from 0 to 0.29, eq. (18) is for Bo > 0.29. The correlations are also presented in Figure 8A for readability. Figure 8B shows that most of the data point lays on the diagonal within a 10% of confidence interval, which means eq. (16), (17), and (18) can adequately describe the trends of Nu dataset. The higher Bo number, the more accurate the correlations are. There is still data scattering in the heat transfer impairment region which is the subject of future work.

## Conclusion

In this work, we discussed and presented 40 DNS simulations of heated parallel plates representing the downcomer of advanced reactors. We investigated a wide range of conditions from forced to natural convection corresponding to a set of Re number from 5000 to 10000 and Gr number from 0 to $10^{10}$. High Pr fluid (FLiBe) has been chosen as working fluid. Validation of appropriate runtime settings has been acquired through comparison with Kasagi DNS data and then applied to all simulation cases. The time averaged turbulence statistics of velocity and temperature for each simulation case have been calculated. The calculation results show that there is a strong impact in THF budgets as the Ri number increases, where the enhancement the buoyancy force on the hot and cold wall side occurs. Changes in Pr number dramatically affect the flow behavior. In particular, the amplitude of turbulence statistics decreases as the Pr number grows, and the bigger Ri is, the stronger this effect will become. In some conditions, the Pr effect can reduce the mixed convection behavior, making mixed convection cases behave nearly like a force convection. At the same Pr number but different Re and Ri number, mix convection cases depend largely on the Ri number, whereas the Re number alone has a small impact on the flow behaviors. Finally, we observed a peculiar budget peaking in the axial direction, which has been investigated by implementing POD.

We also collected time series and Nusselt number data. The PSD results show that highest PSD value are observed in the low-frequency region. Moreover, we observe an energy shift from low to high frequency as buoyancy becomes more dominant: the shift at high frequency occurs at around 20 and 70 Hz. We also observe a temperature PSD peak for case 1: suggesting the presence of coherent structures.

The assembled Nusselt number dataset shows a good overall agreement with experimental data and available correlations. The behavior of Nu number in the heat transfer deterioration region is in good agreement with what is observed in budgets calculation. A generalized correlation for Nu data in fully developed regions has also been proposed to describe this trend, and the correlation fits most of experimental and DNS data within a 10% of confidence interval. In the future, we aim to add more experimental and DNS data to our assessment to confirm and justify the correctness of the proposed correlation.

## Acknowledgements

This work was funded by a U.S. Department of Energy Integrated Research Project entitled "Center of Excellence for Thermal-Fluids Applications in Nuclear Energy: Establishing the knowledgebase for thermal-hydraulic multiscale simulation to accelerate the deployment of advanced reactors -IRP-NEAMS-1.1: Thermal-Fluids Applications in Nuclear Energy. The computational resources are provided by the Oak Ridge Leadership Computing Facility at the Oak Ridge National Laboratory, which is supported by the Office of Science of the U.S. Department of Energy under Contract No. DE-AC05-00OR22725.